\documentclass[aps,prl,reprint,groupedaddress]{revtex4-1}
\usepackage[utf8]{inputenc}
\usepackage{amsmath}
\usepackage{amsfonts}
\usepackage{amssymb}
\usepackage{graphicx}
\usepackage{upgreek}
\begin{document}

\title{New measurement of the $1S-3S$ transition frequency of hydrogen: 
contribution to the proton charge radius puzzle}

\author{Hélène Fleurbaey}
\author{Sandrine Galtier}
\altaffiliation{Present address: Institut Lumière Matière, 
UMR 5306 Université Lyon 1-CNRS, Université de Lyon, 69622 Villeurbanne cedex,
France}
\author{Simon Thomas}
\author{Marie Bonnaud}
\author{Lucile Julien}
\author{François Biraben}
\author{François Nez}
\affiliation{Laboratoire Kastler Brossel, Sorbonne Université, CNRS, ENS-Université PSL, Collège de France, 4 place Jussieu, Case 74,
75252 Paris Cedex 05, France}

\author{Michel Abgrall}
\author{Jocelyne Guéna}
\affiliation{LNE-SYRTE, Observatoire de Paris, ENS-Université PSL, CNRS, 
Sorbonne Université, 
61 avenue de l'Observatoire, 75014 Paris, France}
\date{\today}

\begin{abstract}
We present a new measurement of the $1S-3S$ two-photon transition frequency of 
hydrogen, realized with a continuous-wave excitation laser at 205 nm on a 
room-temperature atomic beam, 
with a relative uncertainty of $9\times10^{-13}$.
The proton charge radius deduced from this 
measurement, $r_\text{p}=0.877(13)$~fm, 
is in very good agreement with the current CODATA-recommended value.
This result contributes to the ongoing search to solve the proton charge radius 
puzzle, which arose from a discrepancy between the CODATA value and a 
more precise determination of $r_\text{p}$ from muonic hydrogen spectroscopy.
\end{abstract}


\maketitle

Hydrogen is a cornerstone of atomic physics, as it plays a key role in the 
determination of the Rydberg constant and in testing fundamental theories such 
as the quantum electrodynamics (QED) theory.
Since it is the simplest atom, the energy levels of hydrogen are described 
theoretically with a good accuracy, and can be written as the sum of two terms. 
The first term is directly linked to the Rydberg constant $R_\infty$. It takes 
into account the solution of the Dirac equation 
and the leading-order recoil correction due to the finite mass of the proton.
The second term, commonly known as the Lamb shift, includes QED and 
relativistic contributions, as well as the finite nuclear size effect 
characterized by the proton rms charge radius $r_\text{p}$.
Hydrogen spectroscopy provides an access to differences of energy levels. 
For instance, the $1S-2S$ transition frequency has been 
measured with a relative uncertainty of $4.2\times10^{-15}$ \cite{Parthey11}.
By making an appropriate linear combination of this frequency with that of 
another transition such as the $2S-nS/D$ transitions \cite{deBeauvoir00},
one obtains experimental values of the Rydberg constant and of the ground-state 
Lamb shift, from which the proton radius can be derived assuming that the QED 
calculations are correct.
The global adjustment of fundamental constants realized by the CODATA 
\cite{Mohr16} partly relies on such a scheme, while also including deuterium spectroscopy and
electron-proton/deuteron scattering experimental results \cite{scattering}.

In 2010, the spectroscopy of muonic hydrogen \cite{Pohl10,Antognini13} yielded 
a value of $r_\text{p}$ an order of magnitude more precise, but about 4\% 
smaller, than the CODATA-recommended value. This 
discrepancy has become known as the proton radius puzzle \cite{Carlson15}.
A recent measurement of the hydrogen $2S-4P$ transition frequency in Garching 
\cite{Beyer79} has brought a new dimension to this conundrum, as it agrees with the smaller muonic value of the proton radius,
in disagreement with other spectroscopic measurements in electronic hydrogen.

In this Letter, we present a new measurement of the $1S-3S$ two-photon hydrogen 
transition frequency, realized with a continuous-wave (cw) 205~nm excitation 
laser and detected through the Balmer-$\alpha$ $3S-2P$ fluorescence. For the 
first time, the uncertainty on this transition frequency (2.6~kHz) is 
significantly smaller than the proton radius discrepancy, which corresponds to 
a difference of 7~kHz for the $1S-3S$ transition frequency. 
This result improves previous measurements in Paris \cite{Arnoult10} as well as 
in Garching \cite{Yost1S3S}. At this unprecedented level of precision, it will 
allow comparison with future results from the Garching experiment, which 
measures the same transition with an entirely different setup using a 
picosecond laser excitation.


\begin{figure*}
\includegraphics[width=0.92\textwidth]{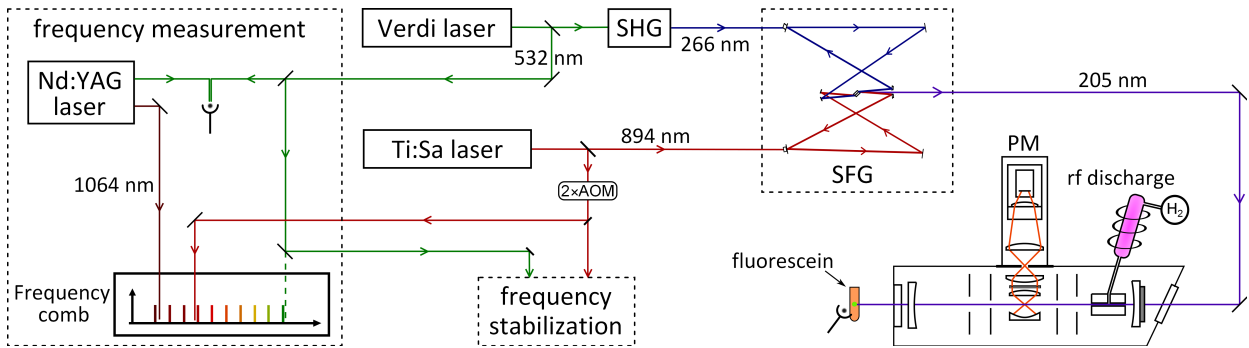}
\caption{\label{setup}Simplified view of the experimental setup. 
Frequency stabilization relies on several Fabry-Perot 
cavities and a Rb-stabilized standard laser. Since 2016, instead of making a 
beatnote directly with the frequency comb, the frequency of the Verdi laser is 
measured via a Nd:YAG transfer laser (more details in text).
SHG: second harmonic generation, SFG: sum frequency generation, 
AOM: acousto-optic modulator, PM: photomultiplier.}
\end{figure*}

The results presented in this Letter were obtained from data recorded in two 
separate sessions, in 2013 \cite{GaltierT} and 2016-2017 \cite{FleurbaeyT}. 
Figure \ref{setup} presents a simplified view of the last version of the experimental setup. 
Our 205~nm cw excitation laser  
 is produced by sum frequency generation (SFG), in a 
$\beta$-barium borate (BBO) crystal, of a home-made tunable titanium:sapphire 
(Ti:Sa) laser at 894~nm and a 266~nm radiation resulting from the frequency 
doubling of a 532~nm laser (Verdi V6 and MBD266, Coherent) \cite{Galtier14}. 
This source delivers between 15~mW (in 2013) and 10~mW (in 2017) at 205~nm, 
depending on the BBO crystal quality and the SFG efficiency.

The frequency stability of the Ti:Sa and Verdi lasers is ensured thanks to  
several Fabry-Perot cavities and a standard laser, a 778~nm laser diode 
stabilized on a two-photon hyperfine transition of $^{85}$Rb \cite{Touahri97,Galtier15}. 
A double-pass acousto-optic modulator (AOM) placed between the Ti:Sa laser and 
the frequency stabilization setup allows to scan the excitation frequency while 
keeping all lasers stabilized.

The Ti:Sa and Verdi laser frequencies, at 894~nm and 532~nm respectively, are measured by comparison with a
MenloSystems femtosecond frequency comb, 
whose 780~nm output is spectrally broadened in a photonic crystal fiber (PCF). 
This frequency comb is referenced 
to the LNE-SYRTE Cs fountain primary frequency standards thanks to a 3-km-long 
optical fiber link \cite{fiberlink}. 
In 2013, the recorded beatnote at 532~nm was weak because of the low power of the frequency comb at this wavelength.
Since 2016, we use an additional laser acting as a transfer 
laser. This cw Nd:YAG laser (Prometheus from Innolight) has two outputs: one at 
532 nm, which is used to make a beatnote with our Verdi laser; the other at 
1064 nm, whose frequency is measured through a beatnote with a new 1064 nm
output of the frequency comb. 

The frequency-stabilized 205~nm laser beam is injected into a power build-up 
cavity, whose axis is collinear with an effusive beam of H atoms formed by the 
dissociation of H$_2$ molecules in a radio-frequency discharge. The cavity 
mirrors have a 25~cm radius of curvature and are placed in a quasi-concentric 
configuration, yielding a waist radius of about 44~$\upmu$m. The Balmer-$\alpha$ 
fluorescence photons are collected 
through a 656~nm interference filter and detected by a photomultiplier.
The entire build-up cavity is inside a vacuum 
chamber, pumped by an oil diffusion pump. A liquid nitrogen trap reduces the oil vapor pressure in the spectroscopy chamber to negligible values. The pressure in the cavity is 
monitored by an ionisation gauge placed on the side of the vacuum chamber, 
which only provides a relative measurement of the actual atomic flux.
The stabilization of the build-up cavity is very 
sensitive to vibrations. In 2015, to improve the signal used for locking, we 
replaced the UV photodiode monitoring the transmitted light with a photodiode 
placed on the side of a quartz tube containing a fluorescein solution. Helmholtz coils, placed around 
the vacuum chamber, create the vertical magnetic field used for velocity 
distribution determination as described below. 

\begin{figure}
\includegraphics[scale=0.48]{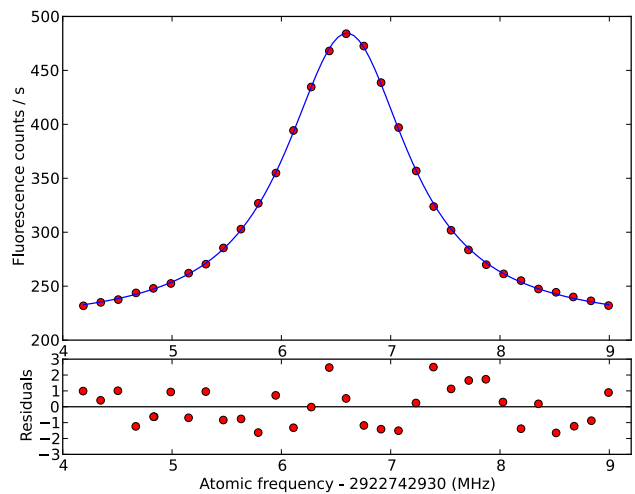}
\caption{\label{signal}Average of 47 recordings of the transition (4-hour 
integration time). No magnetic field was applied. Upper graph: the experimental 
data (red points) is fitted with a theoretical profile (blue line) calculated 
with the velocity distribution parameters $\sigma=1.515$~km/s and 
$v_0=1.23$~km/s (see text). 
The observed linewidth is about 1.35~MHz, as compared to a natural width of 
1~MHz. 
Lower graph: residuals of the fit.}
\end{figure}

To observe the transition, we scan the frequency of the AOM placed in the Ti:Sa 
stabilization loop,
following a predefined back-and-forth 31-point sequence to avoid drifts.
For each AOM frequency point, we record the number of fluorescence photons 
collected by the photomultiplier during one second, as well as the various 
beatnote frequencies. A ``signal'' is obtained by averaging ten such scans.
Figure \ref{signal} shows an average of 47 signals.
We observe a rather large background which is mainly due to UV-induced 
fluorescence of the detection optics.

The main systematic effect in our experiment is the second-order Doppler (SOD) 
effect, which is on the order of 135~kHz and depends on the atomic velocity 
distribution of our room-temperature effusive atomic beam. 
In order to determine this distribution, we follow a method detailed in 
Refs.~\cite{Hagel02,Arnoult10}, in which a vertical magnetic field B is applied in the interaction region, so that an atom moving with velocity 
\textbf{v} experiences a motional electric field 
$\textbf{E}=\textbf{v}\times\textbf{B}$. The Stark shift 
due to this electric field has a quadratic velocity dependence, like the SOD 
shift. At the same time, the Zeeman effect lifts the degeneracy of the $m_F$ hyperfine 
sublevels. The $1S_{1/2}^{F=1}-3S_{1/2}^{F=1}$ transition splits into three 
components, in accordance with the two-photon selection rules ($\Delta m_F=0$). 
The $m_F=0$ component is greatly shifted by the Zeeman effect (about 10~MHz/mT for a magnetic field around 18 mT) and is used to 
calibrate the magnetic field. The two other components are, in first 
approximation, not shifted by the Zeeman effect. 
For a magnetic field of about 18~mT, a level crossing occurs between the 
$3S_{1/2}(F=1,m_F=-1)$ and $3P_{1/2}(F=1,m_F=0)$ levels. The motional Stark 
shift is then large for the $m_F=-1$ component and could compensate the SOD 
shift for this particular sub-transition. 
But the $m_F=\pm 1$ components of the transition are not resolved, since both 
the SOD and Stark shifts are an order of magnitude smaller than the natural 
width of the transition (1 MHz). Thus, the SOD shift is only partly 
compensated. We record the transition signal for no applied magnetic field (residual field of 0.03~mT) and for different values of the 
magnetic field around the level crossing.
To avoid bias due to a possible stray electric field, we also reverse the magnetic field direction. 
Figure~3 shows the apparent line position $\nu_A$, obtained by fitting the line with a simple Lorentzian shape, when the magnetic field is swept around the level crossing.

The analysis relies on a theoretical line profile described 
elsewhere \cite{Arnoult10} which includes the SOD and 
motional Stark shifts. Using the density matrix formalism, it involves 
summing the fluorescence of the $3S(F=1,m_F=0,\pm1)$ sub-levels and that of the 
$3P$ levels to which they can be coupled by the motional Stark effect. The 
$m_F=0$ component only contributes to the signal for a null applied magnetic 
field. The profile is then integrated over a given atomic velocity 
distribution. Our velocity distribution model,
\begin{equation}
f(v, \sigma, v_0)\propto v^3 \text{e}^{-v^2/(2\sigma^2)}P(v/\sigma)\text{e}^
{-v_0/v},
\label{veldist}
\end{equation}
is based on the Maxwellian-type distribution of an effusive beam ($\sigma =\sqrt{kT/M}$, $T$ temperature, $M$ atomic mass) \cite{Ramsey} 
and includes the correction $P(v/\sigma)$ which describes a depletion of slow atoms due to 
interactions within the nozzle \cite{Olander70}. It is multiplied by an 
exponential-decay term to modelize a possible additional depletion of the slow 
atoms in the effusive beam. 
This distribution is fully described by the two parameters $\sigma$ and $v_0$ 
\cite{Galtier15}. 
Moreover, the profile is convoluted with a Lorentzian function to take into 
account broadening effects, mainly due to transit time and pressure broadenings.

\begin{figure}
\includegraphics[scale=0.475]{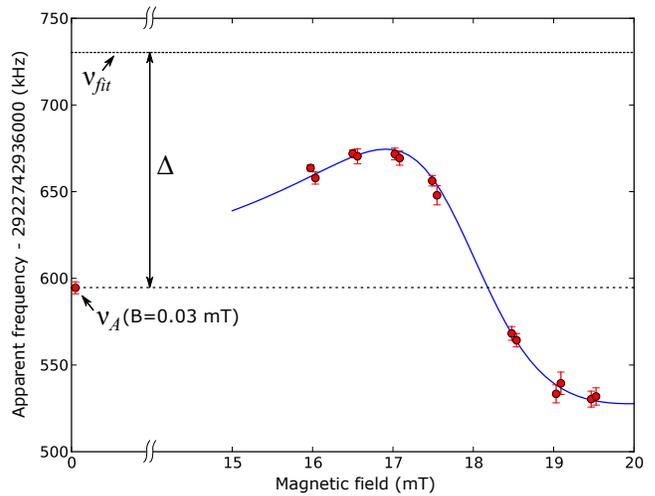}
\caption{\label{dispersion} Experimental (red circles) and calculated (blue curve) apparent positions of the 1S-3S signal as a function of the applied magnetic field B. Each position is given by the center of the best-fitting Lorentzian curve of the experimental or the calculated profile at this B value. The parameters of the velocity distribution used to obtain the blue curve are the ones deduced from the analysis of the LP2 set of data (see below): $\sigma=1.495~$km/s and $v_0= 1.33~$km/s.} 
\end{figure}

The first recording session in 2013, lasting 29 days, yielded 1019
signals recorded for a pressure of $7.5 \times 10^{-5}$ mbar and 7 magnetic field values.
Subsequently, after improving the frequency measurement setup, a second 
recording session was undertaken during 59 days (1700 signals) in 2016-2017. 
This time, the magnetic field procedure was applied for 2 different pressure 
values ($2.7 \times 10^{-5}$ and $2 \times 10^{-4}$ mbar), in order to characterize a possible pressure dependence of the velocity 
distribution. For analysis, we separated the 2016 data in three sets: two sets 
at low pressure (LP1, LP2) recorded before and after the high-pressure set 
(HP). Unfortunately, the pressure gauge was replaced between the two 
recording sessions, so that the indicated pressure values are not comparable between 2013 and 2016-2017.

The four data sets were analyzed independently to determine the velocity 
distribution parameters, through a chi-square minimization process.
Each signal is fitted by theoretical profiles calculated for a grid of 
($\sigma,v_0$) parameters, to determine its center frequency.
The other fit parameters are the amplitude, background offset and Lorentzian 
broadening width.
For a given data set, the mean frequency and the chi-square $\chi^2$ are 
computed. 
The best-fitting velocity distribution parameters are given by the minimum of the 
$\chi^2(\sigma,v_0)$ surface fitted by a polynomial function. 
The results of this minimization for the various data sets are given in the 
first two lines of Table~\ref{values}. Eventually, the signals are fitted 
again using the theoretical profile calculated for the best-fitting velocity 
distribution. The average of this set gives the optimal frequency $\nu_{fit}$ 
which takes into account the SOD, the Zeeman effect and the motional Stark shifts. 

\begin{table}
 \caption{\label{values}Optimal velocity distribution parameters $\sigma$ and $v_0$ and determination of the $1S-3S (F=1)$ frequency. $\nu_A$ is the apparent position of the line for $B=0.03$ mT, $\Delta$ is the difference between the result of the fit procedure $\nu_{fit}$ and $\nu_A$. It corresponds essentially to the SOD (for $B=0.03$ mT the Zeeman shift of the $1S-3S (F=1)$ frequency  is 1.0 kHz) .  $\Delta_{LS}$ is the light shift correction, $\nu_{LS}$ the light shift corrected frequency, $\Delta_{p}$ the pressure correction, $\nu_{LS,p}$ the frequency corrected from the light and pressure shifts, $\Delta_{cd}$ the cross-damping effect. Maser corr. comes from the absolute calibration of the 100 MHz signal used as frequency reference. 
 Only the last four digits of the $1S-3S(F=1)$ frequency are 
 given in the table, $\nu = 2\,922\,742\,936\,$xxx.x~kHz.} 

\begin{ruledtabular}
\begin{tabular}{lrrrr}
Data set &2013 & LP1 &LP2& HP\\
\hline
 $\sigma$ [km/s]  &1.526(27)&1.515(52)&1.495(32)& 1.521(85) \\
$v_0$ [km/s] &0.75(28)&1.23(55)&1.33(31)&0.87(78) \\
\hline
$\nu_A$ [kHz]&592.2(0.7)& 596.8(0.9)&594.4(1.1)&581.6(2.2) \\
$\Delta$  [kHz]  &132.6(1.3)& 137.4(3.8) &135.9(2.1) &131.6(6.8) \\
\hline
$\nu_{fit}$ [kHz] &724.8(1.5)&734.2(3.9)&730.3(2.4)&713.2(7.1) \\
$\Delta_{LS}$ [kHz]&-5.9(1.2)&-10.4(3.0)&-12.1(3.6)&-6.3(10.2) \\
\hline
$\nu_{LS}$ [kHz]&718.9(1.9)&723.8(4.9)&718.2(4.3)&706.9(12.4)\\
$\Delta_{p}$ [kHz]&3.6(2.0)& Pressure               & extrapolation& \\
\hline
$\nu_{LS,p}$[kHz]&722.5(2.8)&                 &722.3(4.9)&                \\
$\Delta_{cd}$ [kHz]&0.6(0.2)&    &0.6(0.2)&   \\
Maser corr. & & & &\\
$ $ [kHz]& -0.599(6)&   &-1.043(6)&       \\
\hline
$\nu_{1S-3S}^{(F=1)}$&722.5(2.8)&  & 721.9(4.9)& \\
\end{tabular}
\end{ruledtabular}
\end{table}

To take into account the light shift, we apply to each signal a frequency 
correction based on a parameter indicating the intra-cavity power (see \cite{SM} for more details). Two such 
parameters have been used: the voltage of the photodiode recording the 
transmitted UV power (for the 2013 recordings) and the square root of the 
two-photon absorption signal height (for 2016-2017). 
As the signal height depends on pressure, the correction coefficient was 
determined separately for each pressure value. The $\Delta_{LS}$ correction
is obtained by a linear extrapolation of the frequency with respect to the chosen 
parameter. 
The light-shift-corrected frequencies $\nu_{LS}$ are given in Table~\ref{values}.

Collisions between atoms can also induce frequency shifts, depending linearly 
on the pressure. 
To determine this pressure shift for the 2013 data set, measurements were 
carried out several times during that recording session, 
for two or three pressure values in the same day, with no applied magnetic 
field. At that time, the velocity distribution was measured for only one 
pressure value and our velocity distribution model could allow for pressure 
dependence of the parameter 
$v_0$, so that we did not know which parameters should be used to analyse the 
other pressure points \cite{Galtier15}. 
The analysis of the 2016 data gave us insight on this question.
In fact, the velocity distribution does not seem to depend significantly on 
pressure, at least within experimental uncertainties (see Table \ref{values}). 
To check this assumption, we have fitted a number of signals using the various 
best-fitting distributions. The resuting change in the center frequency was at most
about 3~kHz.
Hence, when analyzing the 2013 recordings, we use the same velocity 
distribution for all pressure values and we add  in quadrature an uncertainty of 3 kHz
for the points measured at a pressure different from $7.5 \times 10^{-5}$ mbar. 
We thus get a pressure correction of +3.6(2.0)~kHz (see Table \ref{values}). 
For the 2016-2017 session, since the velocity distribution was determined for each 
pressure value, we simply extrapolate the light-shift corrected frequencies of 
the three data sets to zero pressure.
 
At this point, we add a correction of $+0.6(0.2)~$kHz to take into account the 
frequency shift resulting from the cross-damping effect \cite{Hessels10,Hessels11}, following our 
theoretical estimation of this shift \cite{Fleurbaey17}. 

\begin{figure}
\includegraphics[scale=0.475]{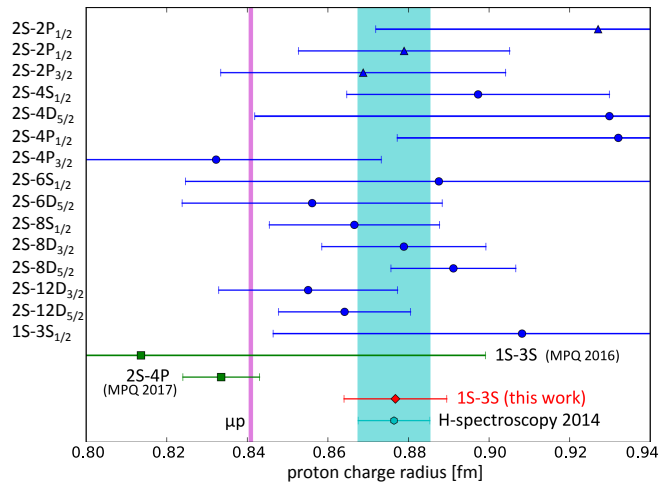}
\caption{\label{radius}Proton charge radius values from H spectroscopy, with 
1$\sigma$ errorbars. The pink bar is the value from muonic hydrogen 
spectroscopy \cite{Antognini13}.  The CODATA-2014 H-spectroscopy average 
\cite{Mohr16} 
(light blue bar and hexagon) includes RF measurements (blue triangles) as well 
as combinations of optical transitions with the $1S-2S$ frequency (blue 
circles). Green squares are obtained from optical transitions measured in 
Garching since 2014 
\cite{Yost1S3S,Beyer79}, and the red diamond is the present work.}
\end{figure}

All the frequency measurements were done with respect to the 100~MHz reference 
signal from LNE-SYRTE. This reference was obtained from a hydrogen maser, whose 
frequency was continuously measured by the LNE-SYRTE atomic fountains realizing 
the frequency of the SI second to a few $10^{-16}$ \cite{Guena,Rovera}. Using a 
simple linear frequency drift of the order of  $10^{-16}$ per day to model the 
H-maser behavior over each period, we estimate the average fractional shift of 
the reference signal with respect to the SI to be $-205(2)\times10^{-15}$ in 
2013, and $-357(2)\times10^{-15}$ in 2016-2017. This yields an absolute 
correction to the $1S-3S$ transition frequency of $-599(6)$~Hz for the 2013 
measurement and $-1043(6)$~Hz for the 2016-2017 measurement. 

The centroid value of the transition is calculated by adding a hyperfine 
correction of $+341\,949.077(3)~$kHz derived from experimental values of the 
$1S$ and $2S$ hyperfine splittings \cite{Karshenboim02}. 
Eventually, we obtain for the two recording sessions,
\begin{eqnarray}
\nu_{1S-3S}^{2013}&=2\,922\,743\,278\,671.6(2.8)~\rm kHz,\\
\nu_{1S-3S}^{2017}&=2\,922\,743\,278\,671.0(4.9)~\rm kHz.
\end{eqnarray}

We estimate a correlation coefficient of 0.186 between the two results.
The weighted average of our two measurements
is then $\nu_{1S-3S}=2\,922\,743\,278\,671.5(2.6)~$kHz. 
Combining this result with the $1S-2S$ transition frequency \cite{Parthey11}, 
one can derive values of the Rydberg constant, 
$R_\infty=10\,973\,731.568\,53(14)~\rm m^{-1}$, and 
the proton charge radius, $r_{\text{p}}=0.877(13)~$fm. 
The latter is shown in Fig.~\ref{radius} along with other determinations 
of the proton radius from hydrogen spectroscopy. The present result
is in very good agreement with
the CODATA-2014 recommended value (0.8751(61)~fm \cite{Mohr16}), and disagrees 
with the value deduced from muonic spectroscopy \cite{Antognini13} by 
$2.8~ \sigma$, thus reinforcing the proton radius puzzle.

In the near future, we plan to cool the hydrogen beam down to the temperature 
of liquid nitrogen, in order to reduce the second-order Doppler shift and 
improve the accuracy of our measurement.

\begin{acknowledgments}
 The authors thank O. Acef for the transfer laser.
This work was supported by the French National Research Agency (ANR) through
the cluster of excellence FIRST-TF (ANR-10-LABX-48), the PROCADIS project 
(ANR-2010-BLANC:04510) and the Equipex REFIMEVE+ (ANR-11-EQPX-0039), and by the 
CNRS.
\end{acknowledgments}

\end{document}